\documentclass{jpsj2}

\title{Pseudogap and Short-Range Antiferromagnetic Correlation 
Controlled Fermi Surface in Underdoped Cuprates:
From Fermi Arc to Electron Pocket}

\author{Takao \textsc{Morinari}\thanks{
morinari@yukawa.kyoto-u.ac.jp
}}

\inst{
Yukawa Institute for Theoretical Physics, Kyoto University, Kyoto 606-8502, Japan
}

\abst{
Motivated by recent quantum oscillation observations in
the underdoped high-temperature superconductors,
the effect of the short-range antiferromagnetic correlations
on the electronic properties of the short-ranged d-density wave state
is investigated.
At intermediate d-density wave correlations,
the cross section of the energy dispersion at the Fermi energy consists of
hole pockets and an electron pocket.
It is argued that the electron pocket feature is smeared out
by the effect of the short-range antiferromagnetic correlation,
which could be the reason why any electron pockets have never been
observed in the angle resolved photoemission experiments.
The Hall resistance is calculated for the system with the electron band
and the hole band using a finite temperature formula
assuming Dirac fermion spectrum for the latter.
It is argued that the scattering effect in the hole band 
should be anomalously large or much suppressed
by superconductivity correlations and/or vortex contributions
to explain the recent quantum oscillation observations.
}

\kword{
high-T$_c$, Fermi arc, antiferromagnetic correlation, quantum oscillation
}

\newcommand{\be}{\begin{equation}}
\newcommand{\ee}{\end{equation}}
\newcommand{\bea}{\begin{eqnarray}}
\newcommand{\eea}{\end{eqnarray}}

\begin{document}
\maketitle

\section{Introduction}
Recent quantum oscillation observations in the longitudinal 
and Hall resistivities under high-magnetic fields 
\cite{DoironLeyraud2007,LeBoeuf2007,Yelland2008,Bangura2008} 
uncovered the Fermi surface topology
in the underdoped high-temperature superconductors.
From the observed oscillation period, 
it is found that the Fermi surface consists of small pockets.
Further detailed analysis of the sign change of the Hall coefficient,
it was found that the negative Hall coefficient is not associated with
vortices but with the electronic properties.
Namely the small pocket should be an electron pocket.\cite{LeBoeuf2007}

On the origin of the electron pocket, 
several theories have been proposed.
\cite{Millis2007,Harrison2007,Chakravarty2008,Chen2008}
Millis and Norman proposed a scenario based on a stripe order.
\cite{Millis2007}
It is argued that electron pockets should appear
for intermediate stripe correlations.
Chakravarty and Kee (CK) proposed a scenario
for an electron pocket
based on the d-density wave (DDW) state.\cite{Chakravarty2008}
For strong d-density wave order,
the Fermi surface is arc like,
whereas 
at moderate d-density wave order,
an electron pocket appears around $(\pi,0)$.
CK computed the Hall coefficient 
based on a two band model
including both effects of the electron pocket
and the hole pockets using the $T=0$
conductivity formula given by Ando\cite{Ando1974}.

Contrary to the quantum oscillation results, 
any electron pocket has never been observed 
in angle resolved photoemission spectroscopy (ARPES).
In ARPES, an arc-like truncated Fermi surface has been observed
in the underdoped cuprates.\cite{Norman1998}
So the natural question is:
How can we understand the ARPES result and the quantum 
oscillation result consistently?
In this paper, I propose that both experimental results
can be understood in a unified way by
taking into account the finite-range correlation effect
for the antiferromagnetic correlation.
As for the description of the pseudogap state,
the DDW state is assumed.
However, there is no long-range DDW wave order
in the model because the antiferromagnetic correlation length, $\xi_{AF}$,
is assumed to be finite.
(In this sense, the state would be similar to 
the staggered flux correlation proposed by Wen and Lee.
\cite{Wen1996,LeeNagaosaWen2006})

The finite antiferromagnetic correlation length effect on the DDW state
is modeled by introducing a Lorentzian distribution
for the wave vectors describing the antiferromagnetic correlation
as in Ref.\citenum{Harrison2007}.
In Ref.\citenum{Harrison2007}, the conventional spin density wave correlation
is assumed.
But here the DDW correlation is assumed because
the ratio of the gap with d-wave symmetry to the 
characteristic temperature for the appearance of the correlation\cite{Won2005}
is consistent with the scaling relation between the pseudogap and the pseudogap 
temperature observed in scanning tunneling spectroscopy.
\cite{Kugler2001}

The condition for the appearance of the electron pocket
is examined and we discuss the effect of finite $\xi_{AF}$.
The electron pocket appears for moderate DDW correlations.\cite{Chakravarty2008}
We show that the electron pocket disappears
for short $\xi_{AF}$.
As for the quantum oscillations,
we calculate the longitudinal resistivity and
the Hall coefficient.
Firstly, we study each band contribution separately 
using the finite temperature formula.
Then, we calculate total resistivity and the Hall coefficient
based on the two band model according to CK.\cite{Chakravarty2008}
We assume the conventional Landau levels for the description 
of the electron pocket while we assume the Landau level wave functions 
for the Dirac fermions for the hole pockets.

The organization of the paper is as follows.
In Sec.\ref{sec_fa},
we introduce the model and discuss the topology of the Fermi surface.
The condition is examined for the appearance of the electron pocket.
We show that the electron pocket feature is smeared out
because of the short-range correlation effect.
In Sec.\ref{sec_sdh},
the longitudinal conductivities are computed 
for the electron pocket and for the hole pocket.
The lifetime broadening effect is discussed.
The total longitudinal resistivity is calculated
based on the two band model.
Section \ref{sec_summary} is devoted to
Summary and Discussion.

\section{Fermi Arc and Electron Pocket}
\label{sec_fa}
Before describing the model to analyze the effect
of short-range antiferromagnetic fluctuations,
we describe the mean field Hamiltonian 
for the DDW state,\cite{Chakravarty2001} or orbital antiferromagnetic 
state.\cite{Nersesyan1989}
The Hamiltonian for the mean field state is 
\bea
H &=& \sum\limits_{{\bf k} \in RBZ} \left( {\begin{array}{*{20}c}
   {c_{{\bf k}\sigma }^\dag  } & {c_{{\bf k} + Q\sigma }^\dag  }  
   \end{array}} \right) \nonumber \\ 
& & \times
   \left( {\begin{array}{*{20}c}
   {\varepsilon _{\bf k}  - \mu } & {i\Delta _{\bf k} }  \\
   { - i\Delta _{\bf k} } & {\varepsilon _{{\bf k} + {\bf Q}}  - \mu }  \\
\end{array}} \right)\left( {\begin{array}{*{20}c}
   {c_{{\bf k}\sigma } }  \\
   {c_{{\bf k} + {\bf Q}\sigma } }  \\
\end{array}} \right),
\label{eq_H}
\eea
where ${\bf Q}=(\pi,\pi)$ and
\bea
\varepsilon _{\bf k}  &=&  - 2t_0 \left( {\cos k_x  + \cos k_y } \right) 
- 4t_1 \cos k_x \cos k_y  \nonumber \\
& & - 2t_2 \left( {\cos 2k_x  + \cos 2k_y } \right).
\eea
The lattice constant, $a$, is set to be unity except in 
making comparisons with other length scales.
The summation with respect to ${\bf k}$ is taken over the
reduced magnetic Brillouin zone defined by
by $|k_x-k_y|<\pi$ and $|k_x+k_y|<\pi$.
The DDW correlations are described by
\be
\Delta _{\bf k}  = \frac{{\Delta _0 }}{2}\left( {\cos k_x  - \cos k_y } \right).
\ee

The complete analysis of the effect of finite-range fluctuations
requires the self-energy calculation for the Green's function
with the number of components being 
proportional to the system size.
This is a formidable task.
So we introduce several approximations.
First of all, we take the fluctuations are quasi-static.
Since there is no antiferromagnetic long-range order,
the fluctuations should be dynamical.
The time scale is given by the inverse of the spin wave excitation gap.
However, if the doping concentration is low
then one can expect that the spin wave excitation gap is small.
In that case, 
as pointed out in Ref.\citenum{Harrison2007}
the time scale of the fluctuations can be long
compared to the ARPES experiment time scale 
and the time scale of the cyclotron motion in the 
strong magnetic field,
and such a long time scale behavior of antiferromagnetic domains
is observed by neutron scattering 
experiments in an underdoped high-temperature 
superconductor.\cite{Hayden1991}
Secondly in order to reduce the number of components
of the Green's function, we focus on a pair correlation of electrons
with the wave numbers ${\bf k}$ and ${\bf k}+{\bf q}$.
It is necessary to include at least a pair of these electrons
to incorporate the effect of coherence factors.
From the self-energy calculation in the lowest order 
with respect to the fluctuations,
we find the energy dispersion as
\be
E_{\bf k}^{\left(  \pm  \right)}  = \frac{{\varepsilon _{\bf k}  + \varepsilon _{{\bf k} + {\bf q}} }}{2} 
- \mu  \pm \sqrt {\left( {\frac{{\varepsilon _{\bf k}  - \varepsilon _{{\bf k} + {\bf q}} }}{2}} \right)^2  
+ \Delta _{\bf k}^2 },
\label{eq_Ek}
\ee
where ${\bf q}$ dependence is implicit in $E_{\bf k}^{(\pm)}$.
The coupling constant is included through the 
change of the order parameter amplitude, $\Delta_0$.
Anisotropy arising from the fluctuations is
approximately calculated by taking an average over ${\bf q}$
whose distribution is given by
the Lorentzian distribution
with $\xi_{AF}^{-1}$ being broadening factor of the distribution,
\be
\rho ({\bf q})  = \frac{{\xi_{AF}^{ - 1} /\pi }}
{{\left( {{\bf q} - {\bf Q} } \right)^2  + \xi_{AF}^{ - 2} }}.
\label{eq_lorentz}
\ee
In numerical computations for finite size systems, the normalization factor
is rescaled so that $\int d^2 {\bf q} \rho ({\bf q}) = 1$.

\begin{figure}
\centerline{\includegraphics[width=6in]{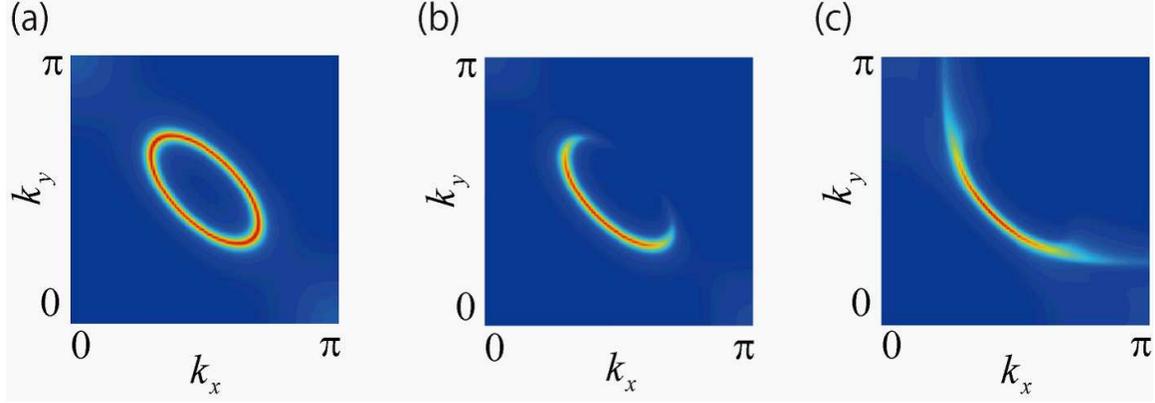}}
 \caption{ \label{fig_fa}
	(Color online) Spectral intensity at the Fermi energy calculated from eq.(\ref{eq_spA}).
	The parameters are $t_1/t_0=-0.25$, $t_2/t_0=0.10$, and $\Delta_0/t_0=1$.
	(a)Spectral intensity without the coherence factors.
	(b)Spectral intensity with the coherence factors.
	(a) and (b) are for the mean field state, ${\bf q}={\bf Q}$.
	(c)Spectral intensity at $\xi_{AF}/a=5$ with $a$ the lattice constant.
    }
 \end{figure}

The spectral function is given by
\bea
 - \frac{1}{\pi }{\mathop{\rm Im}\nolimits} G_{{\bf k}\sigma }^R \left( \omega  \right) 
&=& \frac{{E_{\bf k}  + \varepsilon _{\bf k}^{\left(  -  \right)} }}
{{2E_{\bf k} }}\delta \left( {\omega  - E_{\bf k}^{\left(  +  \right)} } \right) 
\nonumber \\
& & + \frac{{E_{\bf k}  - \varepsilon _{\bf k}^{\left(  -  \right)} }}{{2E_{\bf k} }}\delta 
\left( {\omega  - E_{\bf k}^{\left(  -  \right)} } \right),
\label{eq_spA}
\eea
with $\varepsilon _{\bf k}^{\left(  \pm  \right)}  
= \left( {\varepsilon _{\bf k}  \pm \varepsilon _{{\bf k} + {\bf q}} } \right)/2$,
and $E_{\bf k}  = \sqrt {\varepsilon _{\bf k}^{\left(  -  \right) 2}   + \Delta _{\bf k}^2 }$.
The factors before the delta-functions are the coherence factors
associated with the DDW correlation.
Hereafter the delta-functions in the right hand side
of eq. (\ref{eq_spA}) are replaced by a Lorentzian function,
$(\Gamma/\pi)/\left[ (\omega - E_{\bf k}^{(\pm )} )^2 + \Gamma^2 \right]$
in numerical computations, and set $\Gamma/t_0 = 0.1$.
If we neglect the coherence factors, 
we obtain the so-called hole pockets at ${\bf q}={\bf Q}$ as shown in Fig.\ref{fig_fa}(a)
for $\Delta_0/t_0 = 1$.
As pointed out by Chakravarty {\it et al}.,\cite{Chakravarty2003}
the weight outside the reduced Brillouin zone
is suppressed by including the coherence factors as shown in Fig.\ref{fig_fa}(b).
However, there are some deviations from the ARPES results
around the end points of the arc as pointed out by Norman {\it et al}.\cite{Norman2007}
This discrepancy is removed by
decreasing $\xi_{AF}$ as shown in Fig.\ref{fig_fa}(c).

\begin{figure}
   \centerline{\includegraphics[width=6in]{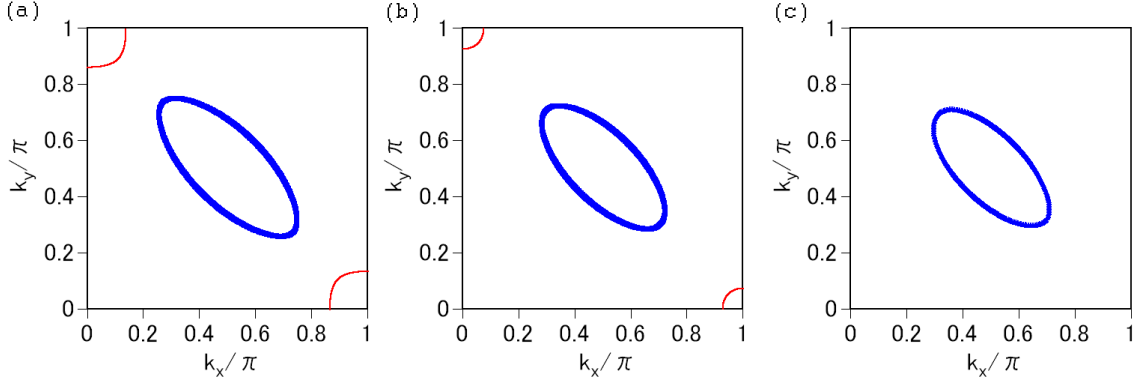}}
   \caption{ \label{fig_ep}
	(Color online) The cross section of the energy dispersion 
	at the Fermi energy 
	in a quadrant of the Brillouin zone:
	(a)$\Delta_0/t_0=0.5$,
	(b)$\Delta_0/t_0=0.7$, and 
	(c)$\Delta_0/t_0=0.8$.
	The hopping parameters are the same as those in Fig.\ref{fig_fa}.
	The thin solid lines around $(\pi,0)$ and $(0,\pi)$
	is associated with $E_{\bf k}^{(+)}=0$, whereas
	the thick solid line around $(\pi/2,\pi/2)$ is
	associated with $E_{\bf k}^{(-)}=0$.
    }
\end{figure}

For $\Delta_0/t_0 \simeq 1$, we only see Fermi arcs.
However, by decreasing $\Delta_0/t_0$ electron pocket appears.
In Fig.\ref{fig_ep}, the cross section of the energy dispersion
at the Fermi energy is shown for different values
of $\Delta_0/t_0$.
For small values of $\Delta_0/t_0$, the electron pocket is clearly seen
around $(\pi,0)$.
The size of the electron pocket decreases with increasing $\Delta_0/t_0$.
The condition for the appearance of the electron pocket is
$E_{{\bf k}=(\pi,0)}^{(+)}=4t_1-4t_2-\mu+\Delta_0 < 0$.
For $x=0.10$ with $t_1/t_0=-0.25$ and $t_2/t_0=0.10$,
this condition is satisfied at $\Delta_0/t_0=0.7$
but not satisfied at $\Delta_0/t_0=0.8$.

\begin{figure}
   \centerline{\includegraphics[width=6in]{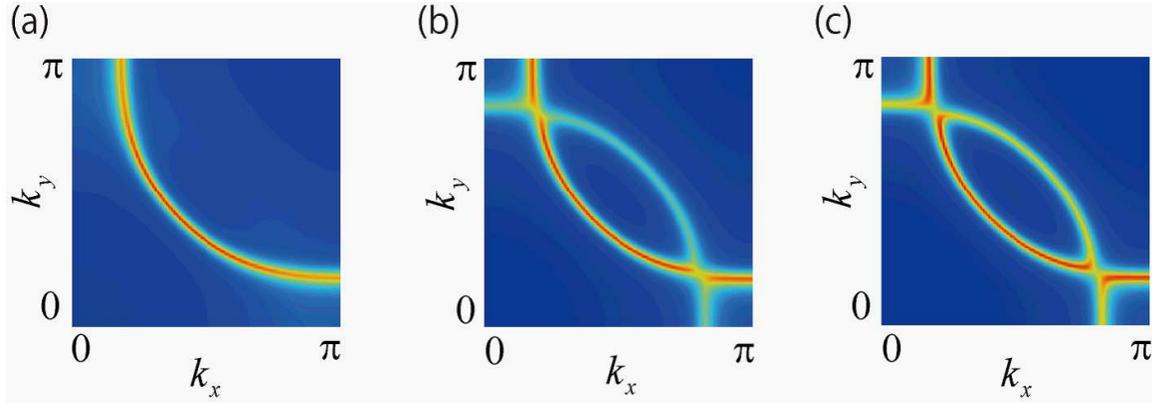}}
   \caption{ \label{fig_ep_xiaf}
	(Color online) Spectral intensity at (a)$\xi_{AF}/a=2$, 
	(b)$\xi_{AF}/a=50$, and (c)$\xi_{AF}/a=100$.
	$\Delta_0/t_0=0.20$ for all panels.
	Coherence factors are turned off to make clear the separation 
	between the electron pocket and the hole pocket.
    }
\end{figure}

\begin{figure}
   \centerline{\includegraphics[width=3.4in]{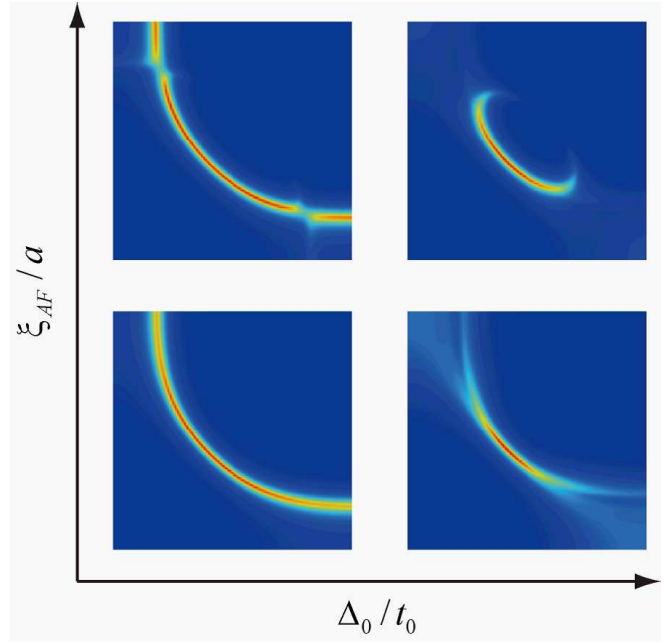}}
   \caption{ \label{fig_summary}
	(Color online) Summary of $\Delta_0/t_0$ and $\xi_{AF}/a$ dependence of the spectral weight.
	For small $\xi_{AF}/a$ and $\Delta_0/t_0$, the full Fermi surface is seen.
	For small $\xi_{AF}/a$ but large $\Delta_0/t_0$, the Fermi surface is truncated,
	and becomes an arc.
	For large $\xi_{AF}/a$ and small $\Delta_0/t_0$, the energy band is
	separated into two bands: One is the hole band and the other is 
	the electron band. 
	The energy bands themselves form closed paths at the Fermi energy.
	For large $\xi_{AF}/a$ and large $\Delta_0/t_0$, 
	there is no electron band, and there are some rounded feature around the end
	points of the arc.
    }
\end{figure}

Although the existence of the electron pocket is consistent with
the recent quantum oscillation observations,
such a pocket has never been observed in ARPES measurements so far.\cite{ShenRMP}
In order to understand this point, 
the spectral weights for several values of $\xi_{AF}$
are shown in Fig.\ref{fig_ep_xiaf}.
For short $\xi_{AF}$, the electron band merges into Fermi arcs and
make a full Fermi surface.
By increasing $\xi_{AF}$, there appears a clear separation 
between the Fermi arc and the electron band.
Thus, the electron pocket disappears for short $\xi_{AF}$.
The qualitative difference of the spectral intensity
at the Fermi energy on $\Delta_0/t_0$ and $\xi_{AF}/a$
plane is summarized in Fig.\ref{fig_summary}.

\section{Shubnikov-de Haas Oscillation}
\label{sec_sdh}
Now we discuss the Shubnikov-de Haas oscillation associated with
the electron pocket.
In the experimental observation by Doiron-Leyraud {\it et al}.,\cite{DoironLeyraud2007}
the quantum oscillation is observed both in 
the longitudinal resistivity and the Hall resistivity.
First we consider the former and next we discuss the latter.
In order to include the finite temperature effect,
we use the following formula for the longitudinal conductivity,
\bea
\sigma _{xx}  &=& \frac{{2\pi e^2 }}{h}\int {dE} 
\left( { - \frac{{\partial f}}{{\partial E}}} \right)\sum\limits_{n = 0}^\infty  
{\left( {n + 1} \right)
\frac{{\hbar \omega _c \Gamma /\pi }}{{\left( {E - E_n } \right)^2  
+ \Gamma ^2 }}} \nonumber \\
& & 
\times \frac{{\hbar \omega _c \Gamma /\pi }}{{\left( {E - E_{n + 1} } \right)^2  
+ \Gamma ^2 }}.
\label{eq_sxx}
\eea
Here spin degeneracy is included and $E_n = (n+1/2) \hbar \omega_c$ are
the Landau level energies.
The derivation is given in Appendix\ref{sec_ap_sxx}.
In the numerical computations, the summation over Landau levels
is taken up to $n=200$.
In this formula, the chemical potential, $\mu$, the constant inelastic scattering
parameter $\Gamma$, and the effective mass $m^*$
are the parameters.

In the experiment, the period of the magnetic oscillation is estimated to be
$\Delta B=530$T.\cite{DoironLeyraud2007}
This implies that the Fermi surface area is $A_k = 0.076\pi^2$.
This value is obtained for the electron pocket
at $\Delta_0/t_0=0.442$
for the parameters of $t_1/t_0=-0.25$ and $t_2/t_0=0.10$.
On the other hand, 
the hole pocket takes $A_{hp}=0.138\pi^2$.
The Luttinger sum rule is satisfied because
there are two hole pockets and one electron
pocket in the reduced Brillouin zone as discussed by CK.\cite{Chakravarty2008}
In this parameter set,
the chemical potential is $\mu/t_0 = -0.750$.
For the electron pocket, the bottom of the band energy
is at $E_{(\pi,0)}^{(+)}=-1.4t_0-\mu+\Delta_0\simeq -0.21t_0$.
This should be equal to $\hbar \omega_c \times \Delta B$.
Assuming $m^*/m=2$ and $\Delta B=530$T, we find
$E_{(\pi,0)}^{(+)}=-360$K and $t_0 \simeq 1700$K.

\begin{figure}
\begin{minipage}{4cm}
  \begin{center}	
   \includegraphics[width=7cm]{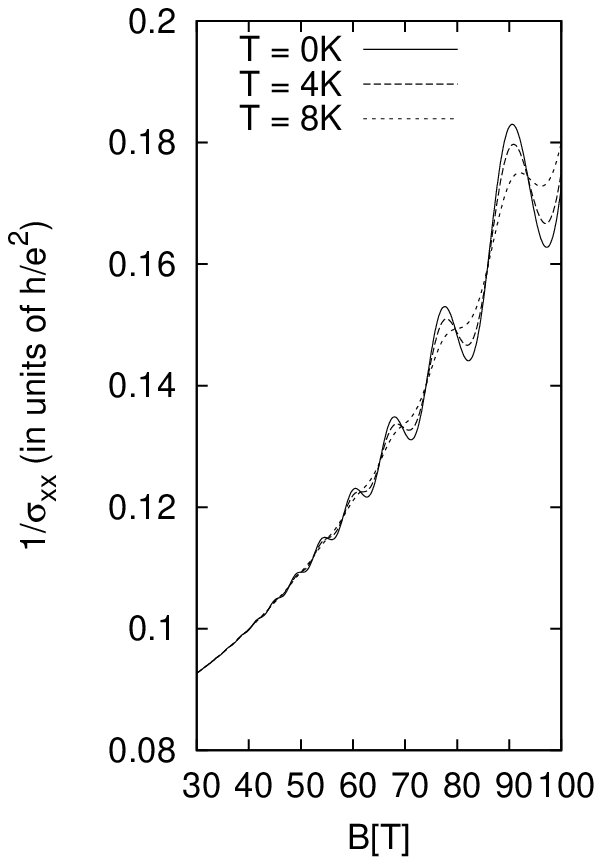}
  \end{center}
\end{minipage}
\begin{minipage}{4cm}
  \begin{center}	
   \includegraphics[width=7cm]{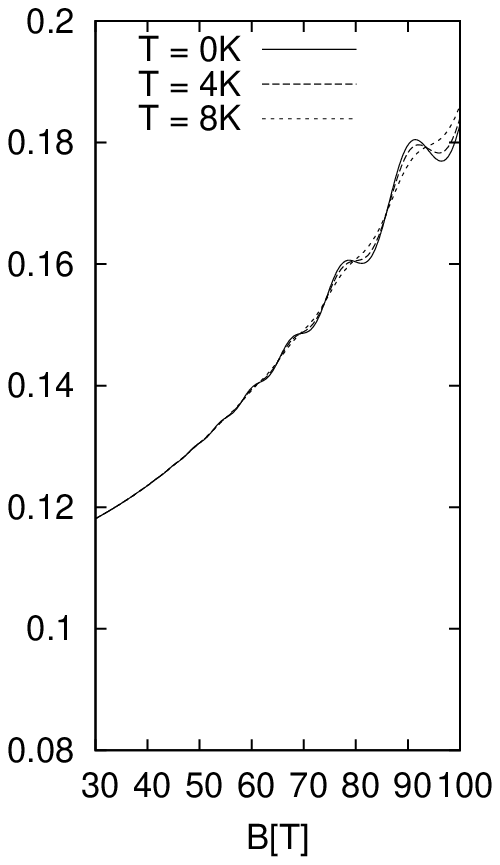}   
  \end{center}
\end{minipage}
\begin{minipage}{4cm}
  \begin{center}	
   \includegraphics[width=7cm]{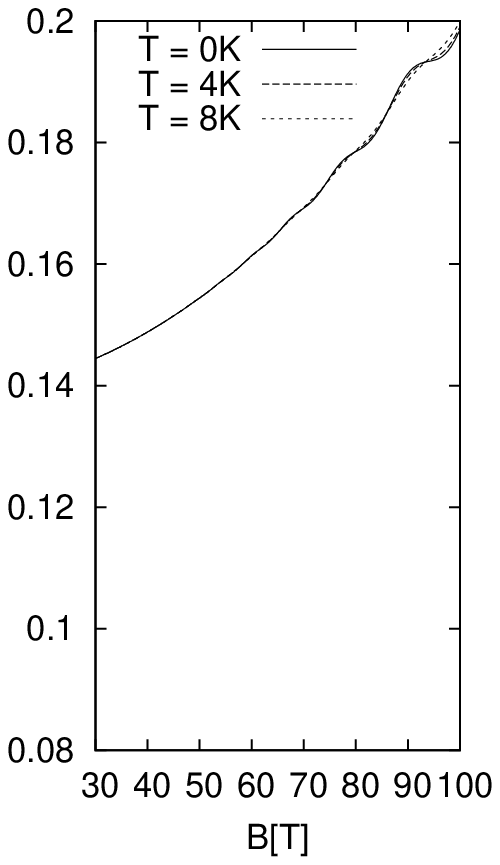}
  \end{center}
\end{minipage}
\caption{ 
	The inverse of $\sigma_{xx}$ for the electron pocket 
	versus magnetic field for different temperatures.
	The left panel is for $\Gamma=30$K, the middle panel is
	for $\Gamma=40$K, and the right panel is for $\Gamma=50$K.
}
\label{fig_ep_sxx}
\end{figure}

In Fig. \ref{fig_ep_sxx} the inverse of $\sigma_{xx}$ 
for the electron pocket is shown.
In the experiments, the oscillation is observed from $B \simeq 50\text{T}$.
To reproduce this behavior $\Gamma$ should satisfy $\Gamma \leq 30\text{K}$
as seen in Fig.\ref{fig_ep_sxx}.
The characteristic magnetic field above which the oscillation
is observed is scaled by $\Gamma$.
This behavior is understood from the fact that
to observe the oscillating behavior 
the separation of the Landau levels 
is larger than the Landau level broadening.

Next we consider the contribution from the hole pocket.
We describe the holes as Dirac particles
because the energy dispersion around $(\pi/2,\pi/2)$ is
well described by a relativistic form.
The necessary parameters 
for the anisotropic dispersion form 
discussed in Ref.\citenum{Nersesyan1989}
are evaluated from the shape of the hole pocket. 
Rotating by $-\pi/4$ about the origin in the $k_x$-$k_y$ plane, the hole pocket 
is described by $\frac{(k_X - \pi/\sqrt{2})^2}{a^2} + 
\frac{k_Y^2}{b^2} =1$.
Here we take the rotated axis as $k_X$ and $k_Y$.
The hole pocket is fitted by taking $a=0.133\pi$ and $b=0.34\pi$.
The approximate energy dispersion is
$E_{\bf k}^{D} = \sqrt {c_X ^2 \left( k_X - \pi/\sqrt{2} \right)^2  
+ c_Y ^2 k_Y ^2 }$.
The parameter $c_X$ is associated with the DDW state
through $c_X=\sqrt{2} \Delta_0$.{\cite{Nersesyan1989}
Setting $\Delta_0=750$K, we find $c_X \simeq 1100\text{K}$
and $c_Y = (0.34\pi/0.133\pi)c_X   \simeq 2700\text{K}$.
The Landau levels are \cite{Nersesyan1989} 
\be
E_n  = \text{sgn}(n) \sqrt {\frac{{2eB}}{{c\hbar }}a^2 c_X c_Y |n|},
\ee
with $n=0,\pm 1, \pm 2, ...$,
where the lattice constant $a$ is recovered.
Substituting above estimated values, we obtain
\be
E_n  \simeq 35\sqrt {nB}.
\label{eq_DiracEn}
\ee
Here $B$ is measured in units of tesla and $E_n$ 
is measured in units of kelvin.
From the value of $A_{hp}=0.138\pi^2$,
the oscillation period associated with the hole pocket
is $970$T from the Onsager relation
as discussed by CK.\cite{Chakravarty2008}
In order to fit with this oscillation period,
the dispersion energy is shifted so that
the chemical potential for the holes is at
$\mu_h \simeq 1100$K.
The formula for the longitudinal conductivity under the magnetic field 
is obtained by 
computing the matrix elements of the Dirac fermion current operators
and taking eq.(\ref{eq_DiracEn}) for $E_n$.
(For explicit derivation and discussions about
impurity scattering effect, see, Ref.\citenum{Shon1998}.)
For degeneracy, we include valley degeneracy as well as spin degeneracy.
In the reduced Brillouin zone, the hole pockets around $(\pi/2,\pi/2)$
and $(-\pi/2,-\pi/2)$ are equivalent.
The same is true to those around $(\pi/2,-\pi/2)$ and $(-\pi/2,\pi/2)$.
However, the hole pockets around $(\pi/2,\pi/2)$ and $(\pi/2,-\pi/2)$
should be distinguished.
Therefore, to incorporate this degeneracy the formula is multiplied 
by the factor of two.
The inverse of $\sigma_{xx}$ for the holes 
is shown in Fig.\ref{fig_hp_sxx} for different $\Gamma$'s.
The behavior is qualitatively similar to that for the electron pocket.
\begin{figure}
\begin{minipage}{4cm}
  \begin{center}	
   \includegraphics[width=7cm]{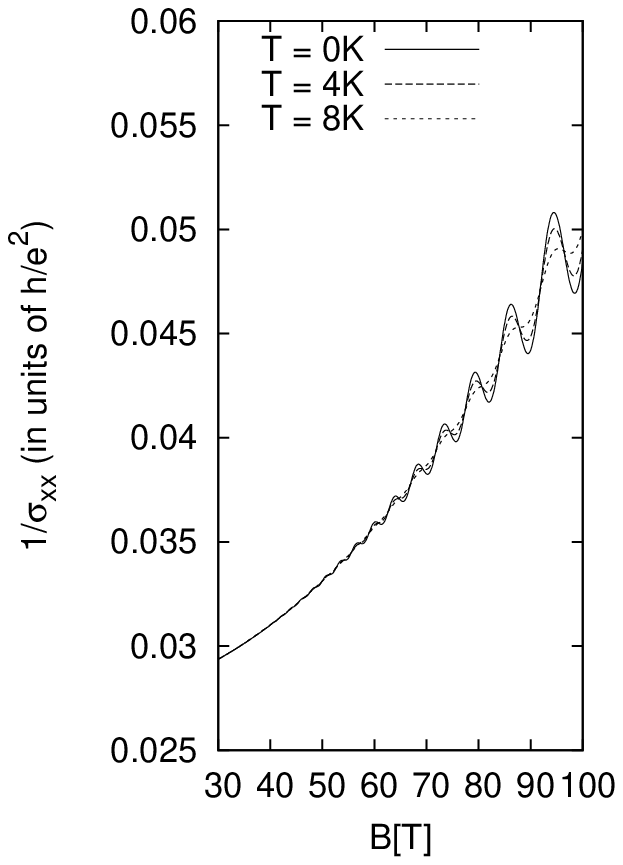}
  \end{center}
\end{minipage}
\begin{minipage}{4cm}
  \begin{center}	
   \includegraphics[width=7cm]{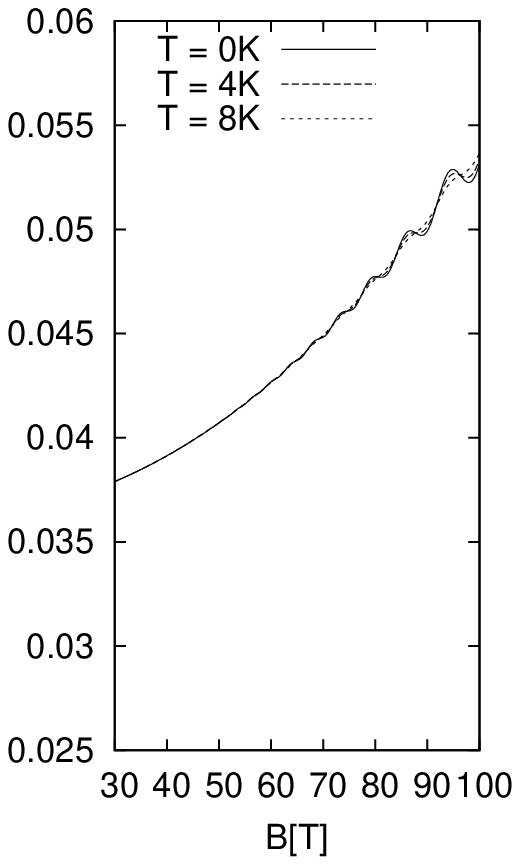}
  \end{center}
\end{minipage}
\begin{minipage}{4cm}
  \begin{center}	
   \includegraphics[width=7cm]{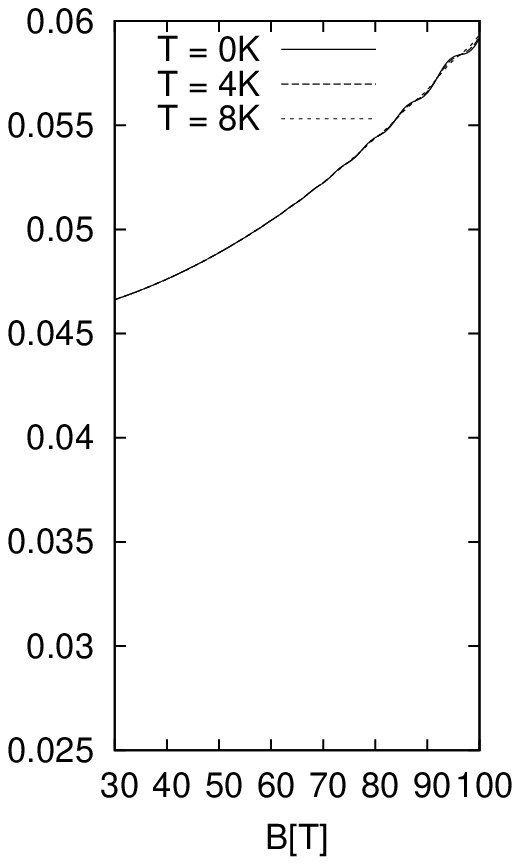}
  \end{center}
\end{minipage}
   \caption{ 
	The inverse of $\sigma_{xx}$ for the hole pocket 
	versus magnetic field for different temperatures.
	The left panel is for $\Gamma=30$K, the middle panel is
	for $\Gamma=40$K, and the right panel is for $\Gamma=50$K.
    }
\label{fig_hp_sxx}
\end{figure}


Now we move on to the total Hall coefficient.
From Fig.\ref{fig_ep_sxx}, we see that $\Gamma$ for the electron band, 
$\Gamma_e$, should be less than $30\text{K}$ to observe the quantum oscillation 
from $B \simeq 50\text{T}$.
Therefore, we fix $\Gamma_e=30\text{K}$ in the following analysis.
The remaining parameter is $\Gamma$ for the hole band, $\Gamma_h$.
We compute the total Hall coefficient
based on the two band formula,
\bea
R_H &=& \frac{{R_H^e \left( {\sigma _{xx}^e } \right)^2  
+ R_H^h \left( {\sigma _{xx}^h } \right)^2  
+ \left( {\sigma _{xx}^e } \right)^2 \left( {\sigma _{xx}^h } \right)^2 
R_H^e R_H^h \left( {R_H^e  + R_H^h } \right)B^2 }}{{\left( {\sigma _{xx}^e  
+ \sigma _{xx}^h } \right)^2  + \left( {\sigma _{xx}^e } \right)^2 
\left( {\sigma _{xx}^h } \right)^2 B^2 \left( {R_H^e  + R_H^h } \right)^2 }},
\eea
assuming $\sigma_{xx}$'s for each band and 
$R_H^{e}=-1/n_e|e|c$ and $R_H^{h}=1/n_h|e|c$
for the Hall coefficients of the electron band and the hole band,
respectively following CK.
Here the particle densities are $n_e=A_{ep}/(2\pi^2)$
and $n_h=A_{hp}/\pi^2$.
The difference from CK is that we use the finite temperature formula
for the computation of the conductivities
and take into account the Dirac fermion properties of the hole band.
From the computation of $R_H$ for various $\Gamma_h$
we find that the behavior of $R_H$ changes largely
depending on $\Gamma_h$.
The total Hall coefficient is shown in Fig.\ref{fig_total_R_H}
for $\Gamma_h=200\text{K}$, 
$\Gamma_h=400\text{K}$, and $\Gamma_h=1000\text{K}$.
For $\Gamma_h=200\text{K}$, $|R_H|$ decreases as increasing $B$.
While for $\Gamma_h=400\text{K}$, $|R_H|$ 
shows an oscillating behavior.
But the amplitude is very small because of the cancellation 
between the electron pocket contribution and the hole pocket contribution.
To reproduce the experimental results,
$\Gamma_h$ should be $\Gamma_h \geq 1000\text{K}$.
Therefore, there should be strong scattering in the hole band.
Although the origin of such a scattering is not clear,
if this is the case the oscillation associated with 
the hole pocket is not observable.
Because we need unrealistically huge magnetic field 
to observe the quantum oscillations associated with the hole pocket
for such large values of $\Gamma_h$.
In Figs.\ref{fig_total_rH} and \ref{fig_total_rxx},
the total Hall resistance and the longitudinal resistivity 
at $\Gamma_h=1000\text{K}$
are shown, respectively.

\begin{figure}
\begin{minipage}{4cm}
  \begin{center}	
   \includegraphics[width=7cm]{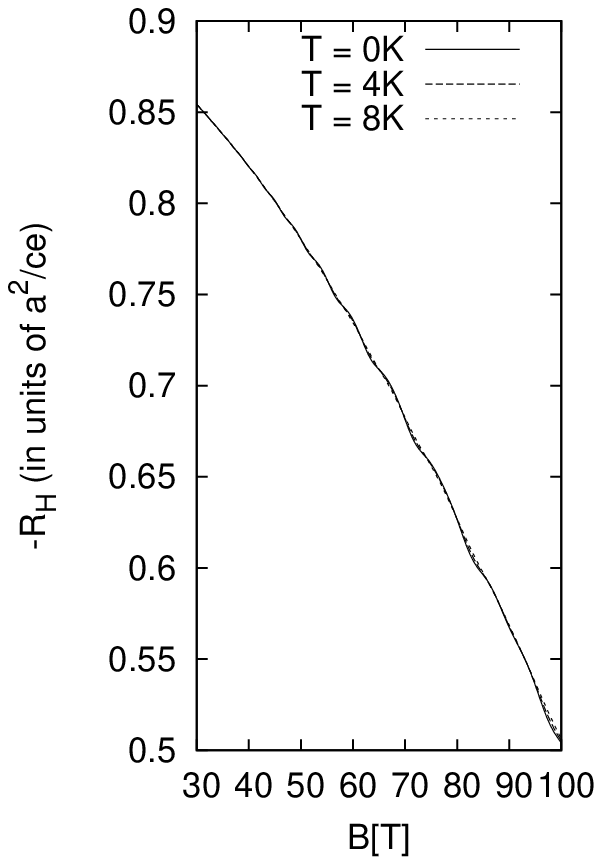}
  \end{center}
\end{minipage}
\begin{minipage}{4cm}
  \begin{center}	
   \includegraphics[width=7cm]{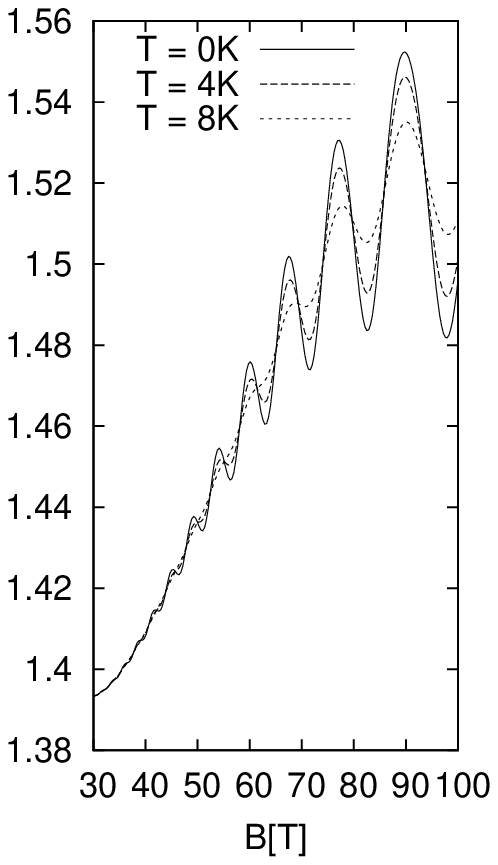}
  \end{center}
\end{minipage}
\begin{minipage}{4cm}
  \begin{center}	
   \includegraphics[width=7cm]{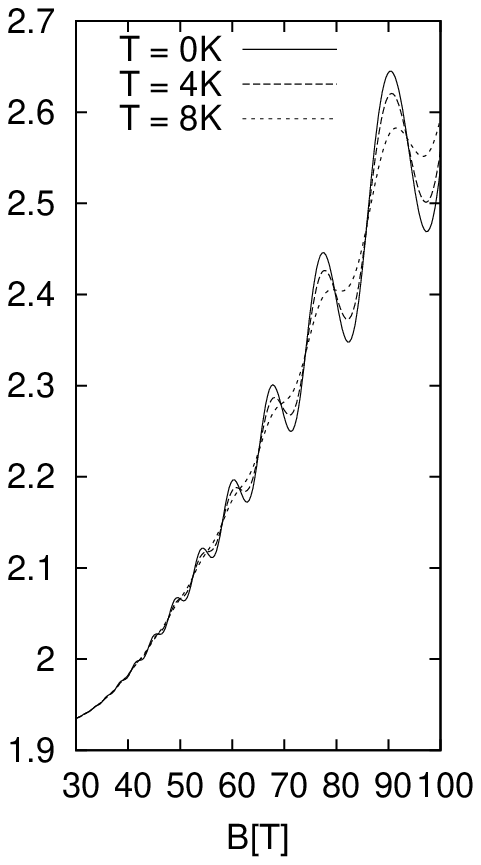}
  \end{center}
\end{minipage}
\caption{ 
	The total Hall coefficient versus magnetic field for
	different temperatures.
	The scattering parameter $\Gamma_h$'s are $\Gamma_h=200\text{K}$ 
	for the left panel,
	$\Gamma_h=400\text{K}$ for the middle panel, 
	and $\Gamma_h=1000\text{K}$ for the right panel.
    }
\label{fig_total_R_H}
\end{figure}

\begin{figure}
   \centerline{\includegraphics[width=3.4in]{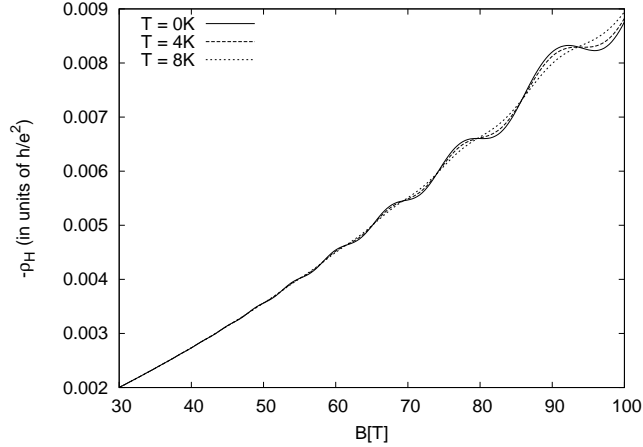}}
   \caption{ \label{fig_total_rH}
	The total Hall resistance versus magnetic field for
	different temperatures.
	The scattering parameters are $\Gamma_e=30\text{K}$
	and $\Gamma_h=1000\text{K}$.
    }
\end{figure}

\begin{figure}
   \centerline{\includegraphics[width=3.4in]{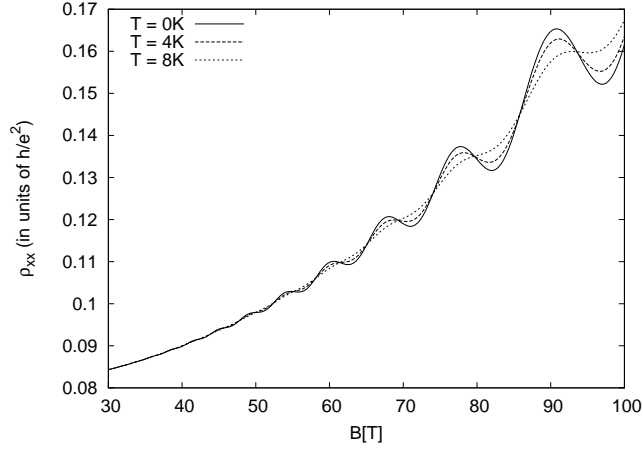}}
   \caption{ \label{fig_total_rxx}
	The total longitudinal resistivity versus magnetic field for
	different temperatures.
	The scattering parameters are $\Gamma_e=30\text{K}$
	and $\Gamma_h=1000\text{K}$.
    }
 \end{figure}

Now we discuss the finite $\xi_{AF}$ effect on the quantum oscillation.
The distribution of the electron pocket and the hole pocket areas 
in the wave vector space is computed by taking the average over
the Lorentz distribution function with respect to ${\bf q}$.
The same analysis is carried out for the hole pocket
in the case of the spin density wave state by Harrison {\it et al}.
\cite{Harrison2007}
The result is well fitted by the Lorentz function.
\begin{figure}
\centerline{\includegraphics[width=3.4in]{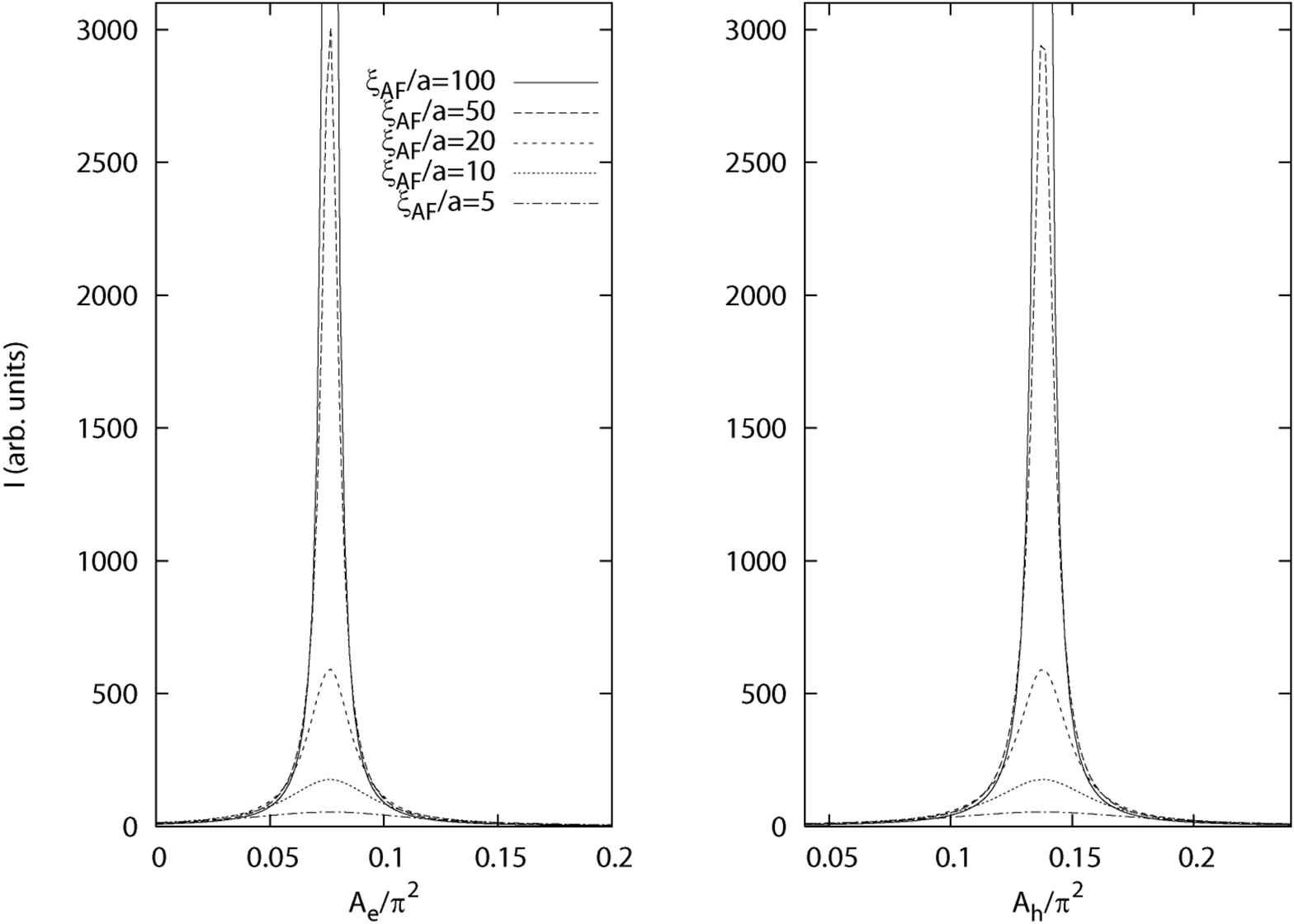}}
\caption{ \label{fig_ak_dist}
	Probability distributions of the electron pocket and hole pocket areas 
	$A_e$ and $A_h$ in the wave vector space for 
	$\xi_{AF}/a = 100, 50, 20, 10, 5$.
}
\end{figure}
This Lorentz form suggests that the oscillating component amplitude
decays rapidly in the low magnetic field
as $\exp \left( - 2\pi \delta B_0/B \right)$
with $\delta B_0 = 7\times 10^3 \delta A_k/\pi^2$.
Here $\delta A_k$ is the half value of width of the Lorentz distribution.
The Lorentz function fits for the electron pocket 
and the hole pocket distributions are
shown in Fig.\ref{fig_ak_dist} for different values of $\xi_{AF}/a$.
We see that at $B=60\text{T}$ the oscillation amplitude is
reduced to $20\%$ at $\xi_{AF}/a=100$ and
$4\%$ at $\xi_{AF}/a=50$.
This analysis suggests that to observe the quantum oscillation
we need sufficiently long $\xi_{AF}$.
This point is first emphasized by Harrison {\it et al}\cite{Harrison2007}
from the analysis of the hole pocket.

\section{Summary and Discussion}
\label{sec_summary}
In this paper, we have studied the effect of the short-range antiferromagnetic
correlation on the short-ranged DDW state.
For $\Delta_0/t_0 \sim 1$, the Fermi surface is arc-like
and there is no electron pocket.
For moderate values of $\Delta_0/t_0$, an electron pocket
appears.
The effect of the short-range antiferromagnetic correlation
on the electron pocket depends on the value of $\xi_{AF}/a$.
For $\xi_{AF}/a \sim O(1)$, the electron pocket 
is smeared out and is combined with the Fermi arc
so that the resulting Fermi surface is the full Fermi surface.
While for $\xi_{AF}/a \gg 1$, the electron pocket 
feature is preserved.
We apply this observation to the recent quantum oscillation 
result and the ARPES results.
The presence of an electron pocket is suggested from the recent
quantum oscillation measurements
\cite{DoironLeyraud2007,LeBoeuf2007,Yelland2008,Bangura2008}
in high magnetic fields.
By contrast, any electron pocket has never been observed in ARPES.
This apparent contradiction is resolved if we assume that
in high magnetic fields $\Delta_0/t_0$ decreases
and $\xi_{AF}/a$ increases.

As argued by Harrison {\it et al}., 
long $\xi_{AF}$ in high magnetic fields 
is consistent with the neutron scattering experiments
and $\xi_{AF}$ can be long at high magnetic fields
because of enhancement of antiferromagnetic correlations
around vortices.
However, the magnetic field effect on the value of $\Delta_0/t_0$
is not well understood.
As demonstrated by Nguyen and Chakravarty\cite{Nguyen2002}
within the mean field theory, 
the magnetic field effect on $\Delta_0$ is negligible.
Therefore, to reduce the value $\Delta_0/t_0$
we need to increase $t_0$.
From the study of the single hole system by
a self-consistent Born approximation in the slave-fermion theory
of the t-J model,\cite{KaneLeeRead1989}
it is believed that the band width is strongly renormalized 
from the bare band width to the order of $J$,
which is consistent with the ARPES results 
in the undoped cuprates.\cite{Wells1995}
One way to suppress the renormalization effect by magnetic field
is to include the induced parallel spin configuration effects.
If we use the susceptibility of the Heisenberg antiferromagnet
at zero temperature,\cite{Chakravarty1988}
the induced hopping amplitude due to the parallel alignment component of 
neighboring spins is only $10\text{K}$ at $B=60\text{T}$.
Although we may expect that this value increases for doped compounds, 
the estimation requires the analysis of the spin disordering effect.
Exact diagonalizations of the $t-t'-t''-J$ model suggest 
that $t_0$ should increase by suppressing frustration effects.\cite{Shibata1999}

In order to examine the effect of scattering 
and the effect of the short-range antiferromagnetic
correlation on the quantum oscillations,
we have calculated the longitudinal conductivities
of the electron band and the hole band.
The electron band is associated with the electron pocket
and the hole band is associated with the hole pocket.
This splitting of the band takes place due to the antiferromagnetic
correlations.
We describe the Landau levels of the electron band
as the conventional non-relativistic Landau levels.
While we describe the hole band Landau levels 
as the Dirac fermion Landau levels.
The parameters of the Dirac fermion dispersion 
are determined from the shape of the hole pocket.
We have computed the longitudinal conductivity 
for each band.
To observe the quantum oscillations from $B \simeq 50\text{T}$, 
the scattering parameter $\Gamma$ should be less than $30\text{K}$.
The total Hall coefficient and the longitudinal resistivity
are computed by the two band formula for the electron band
and the hole band.
It is found that the amplitude of quantum oscillations
is rapidly suppressed by decreasing $\xi_{AF}$.
In order to observe quantum oscillations, we need sufficiently 
long $\xi_{AF}$.
For $\xi_{AF}/a = 100$, the amplitude is reduced to $20\%$
and for $\xi_{AF}/a = 50$, the amplitude is reduced to $4\%$.

In addition we have found that to fit the experiments
we need large $\Gamma$ for the hole band.
The origin of this strong scattering is not associated with
the short-range antiferromagnetic correlation effect.
From the calculation of the spectral weight
we see that there is broadening effect associated with 
the short-range antiferromagnetic correlation effect.
However, the broadening is stronger for the electron band
than for the hole band.
Another possibility for the suppression of the hole band 
contribution is the superconductivity correlation
and the effect of vortices.
If the density of states of the hole band is reduced by
superconductivity, then the hole band contribution is suppressed.
In addition, vortices contribute to the longitudinal resistivity.
This point requires further investigations.

In a recent experiment, de Haas-van Alphen effect is observed
using a magnetic torque technique and
a new oscillation period is discovered.\cite{Sebastian2008}
Podolsky and Kee proposed that the ortho-II potential in 
YBa$_2$Cu$_3$O$_{7-\delta}$ can lead to Fermi pockets
one of which is consistent with the experiment.
However, in the experiment the amplitude is much smaller than that
associated with the electron pocket.
In addition, the effective mass estimated from the newly found oscillation
is twice as large as the effective mass estimated from the oscillation
associated with the electron pocket.
As for the origin of the new oscillation period, more experimental and theoretical
studies would be required.

\section*{Acknowledgment}
I would like to thank Profs. T. Tohyama and K. Maki for helpful discussions.
This work was supported by the Grant-in-Aid for the Global COE Program 
"The Next Generation of Physics, Spun from Universality and Emergence" 
from the Ministry of Education, Culture, Sports, Science and Technology (MEXT) 
of Japan.
and Yukawa International Program for Quark-Hadron Sciences at YITP.
The numerical calculations were carried out in part 
on Altix3700 BX2 at YITP in Kyoto University.

\appendix
\section{Derivation of Eq.(\ref{eq_sxx})}
\label{sec_ap_sxx}
In this appendix, we derive the formula eq.(\ref{eq_sxx}).
From the Kubo formula, the longitudinal conductivity is given by
\be
\sigma_{xx} \left( \omega \right)
=\frac{1}{i\omega} 
\left[
K(\omega ) - K(0)
\right],
\ee
where
\be
K(\omega ) = -\frac{1}{i\hbar S}
\int_0^{\infty} dt \text{e}^{i\omega t-\delta t}
\langle \left[ 
J_x (t), J_x
\right] \rangle,
\ee
with $S$ the area of the system and $\delta$ being an infinitesimal 
positive number.
Here the current operator is
\be
J_x = -\frac{e}{m} \sum_j p_{jx},
\ee
with $p_{jx}$ the single particle momentum operator in the $x$-direction.
If we neglect the interaction between the electrons,
the formula is rewritten in terms of the one-body quantum states.
Taking the $\omega \rightarrow 0$ limit and noting $K(0)=ne^2/m$,
the dc conductivity is
\bea
\sigma_{xx} &=& \frac{2\pi \hbar}{S} \left( \frac{e}{m} \right)^2
\int dE \left( -\frac{\partial f}{\partial E} \right) \nonumber \\
& & \times \sum_{\alpha} 
\langle \alpha |
\delta(E-H) p_x 
\delta(E-H) p_x | \alpha \rangle.
\label{eq_sxxa}
\eea
Here $H$ is the one-body Hamiltonian, $\alpha$ denotes quantum states,
and spin degeneracy is included.
In the absence of the impurity scattering, the delta function part is
\be
\delta(E-H) = \frac{1}{2\pi i} \left( 
\frac{1}{E-H-i\delta} - 
\frac{1}{E-H+i\delta} \right).
\ee
We include the scattering effect by rewriting 
the right hand side as
\be
\frac{1}{2\pi i} \left( 
\frac{1}{E-H-i\Gamma} - 
\frac{1}{E-H+i\Gamma} \right).
\ee

Now we include the magnetic field effect.
Under strong magnetic field, the single body quantum state
is quantized into Landau levels.
Taking the Landau gauge ${\bf A}=(0,Bx,0)$, the wave function
for the $n$-th Landau level with $X$ the center coordinates
is given by
\be
\phi_{nX} (x) = \frac{1}{\sqrt{2^n n! \pi^{1/2} \ell_B}}
H_n \left( \frac{x-X}{\ell_B} \right)
\exp \left( -\frac{(x-X)^2}{2\ell_B^2} \right),
\ee
where $H_n(x)$ are the Hermite polynomials and
$\ell_B=\sqrt{c\hbar/eB}$ is the magnetic length.
The quantum states $|\alpha \rangle$ in eq.(\ref{eq_sxxa})
are now $|n,X \rangle$.
In terms of these eigen-states, the matrix element of the 
momentum operator is
\bea
\langle n', X'| p_x | n,X \rangle
&=& \frac{i\hbar}{\ell_B} \delta_{X'X} \nonumber \\
& & \times \left(
\sqrt{\frac{n+1}{2}}  \delta_{n',n+1}
- \sqrt{\frac{n}{2}}  \delta_{n',n-1}
\right).
\eea
Substituting this formula into eq.(\ref{eq_sxxa}), we obtain
eq. (\ref{eq_sxx}).
If the impurity scattering effect is included 
in the self-consistent Born approximation,
$\Gamma \propto \sqrt{B}$ as shown 
in Ref.\citenum{Ando1974}.
We do not include this effect for simplicity.

\bibliography{../../references/tm_library2}

\end{document}